\documentclass{aa}

\usepackage{times}
\usepackage{amsmath}
\usepackage{epsfig}

\begin{document}

   \thesaurus{11     (11.04.1;11.19.2;12.04.3) }

\title{Theoretical aspects of the inverse Tully-Fisher relation as a
distance indicator: incompleteness in $\log{V_{\rm Max}}$, the 
relevant slope, and the calibrator sample bias}

   \author{P. Teerikorpi\inst{1}, T. Ekholm\inst{1}, M. O. Hanski\inst{1},
	G. Theureau\inst{2,3}}

   \offprints{ }

   \institute{Tuorla Observatory, 21500 Piikki\"{o}, Finland
         \and
	 Osservatorio di Capodimonte, Via Moiariello 16, 80131 Napoli, Italy
	 \and
	 Observatoire de Paris-Meudon, CNRS URA1757, F92195 Meudon 
	 Principal Cedex, France 
             }

   \date{Received / Accepted }

\authorrunning{P. Teerikorpi et al.}
\titlerunning{Inverse Tully-Fisher relation}

   \maketitle

   \begin{abstract}

We study the influence of the assumption behind the
use of the inverse Tully-Fisher relation:\ that there should be no
observational cutoffs in the TF parameter $\log{V_{\rm M}}$.  It is noted
how lower and upper cutoffs would be seen in a $\log{H_0}$ vs.\
``normalized distance'' diagram.  Analytical expressions,
under the simplifying assumption of a normal distribution and the use
of the correct TF slope, are derived for the
resulting biases, especially the average bias which $\log{V_{\rm M}}$ cutoffs
produce in the derived value of $H_0$.
This bias is shown to be relatively weak, and as such
cannot explain the large differences in the reported values of $H_0$ 
derived from direct and inverse TF relations.

Some problems of slope and calibration are shown to be more serious.  
In particular, one consequence of
fitting through the calibrators either the slope relevant for
field galaxies or the steeper slope followed by calibrators is that the derived
value of the Hubble constant comes to depend on the nature of
the calibrator sample. If the calibrator sample is not representative
of the cosmic distribution of $\log{V_{\rm M}}$, large errors in the
derived value of $H_0$ are possible. Analytical expressions are given
for this error that we term the calibrator sample bias.


\end{abstract}

%Section 1
\section{Introduction}

   There have been sincere hopes that the inverse
Tully-Fisher (TF) relation could overcome the distance dependent
selection bias, the so-called Malmquist bias of the 2nd
kind (Teerikorpi \cite{teer97}), influencing the 
determination of the value of the Hubble constant.  
First, Schechter (\cite{sche80})
pointed out that the inverse relation of the form 
\begin{equation} % 1
p = a'M + b' 
\end{equation} % 1
where $M$ is absolute magnitude (or linear diameter) and $p$ is an
observable parameter not affected by selection effects (e.g.\ the
maximum rotation velocity $\log{V_{\rm M}}$ in the TF relation), can be
derived even from magnitude limited samples in an unbiased
manner.  Secondly, Teerikorpi (1984, or \cite{T84}) showed explicitly that
the inverse relation gives unbiased average distances for a
sample of galaxies (if $p$ is not affected by selection), a result later
confirmed by Tully (\cite{tull88}) via simulations (see also
Hendry \& Simmons \cite{hend94}).  Thirdly, Ekholm \&
Teerikorpi (1997, or \cite{ET97}) showed that with the inverse
relation, one does not necessarily need a volume-limited
calibrator sample, which on the contrary is quite critical for
the use of the direct relation:
\begin{equation} % 2
M = a p + b
\end{equation} % 2
   The direct TF relation has been succesfully applied for derivation of $H_0$
by the method of normalized distances that was first used by
Bottinelli et al.\ (\cite{bott86}). Theureau et al.\ (\cite{theub97}) 
have used an improved version of this method in an analysis of a sample of
5171 spiral galaxies that leads to the value of $H_0 \approx 55 \pm 5
\mathrm{\,km\,s^{-1}\,Mpc^{-1}}$.  In this method, which is rather
similar to the Spaenhauer diagram method of Sandage (\cite{sand94}),
one identifies the so-called unbiased plateau in the $\langle \log{H} \rangle$
vs.\ normalized distance diagram for a magnitude or angular diameter
limited sample, from which $H_0$ is determined without the need to
make further model dependent correction due to selection. Normally, 
the unbiased 
plateau contains only 10 - 20 percent of the total sample.  That one
can use the whole sample without bias correction is a
remarkable advantage of the inverse relation approach.

   However, several problems balance the advantages of the
inverse relation (Fouqu\'{e} et al.\ \cite{fouq90}, Teerikorpi 
\cite{teer90}, Sandage et al.\ \cite{sand95}, \cite{ET97}), and
one has not yet been able to use it reliably for an independent
determination of $H_0$ (see Theureau \cite{theu97}, Ekholm
et al.\ \cite{ekho98}).  In the present
paper, we study two important points which must be clarified
before a successful use of the inverse TF relation:
\begin{enumerate}
\item  How would possible cutoffs in $\log{V_{\rm M}}$ be seen in the data?
\item How large an influence do upper and lower cutoffs in
$\log{V_{\rm M}}$ have on derived average distances (and hence on $H_0$)? 
\item  How does the nature of the calibrator sample influence the derived
average distances if the inverse TF slopes for calibrators and
field galaxies differ?
\end{enumerate}

As regards the first point, we emphasize that our approach follows the lines
which we have adopted and found useful with the direct TF relation:\
to detect and overcome the bias without detailed modeling, by studying
how the cutoffs would appear in kinematical applications (c.f.\ the
increase of $\langle H \rangle$ as a function of true 
(or normalized) distance, when the direct
TF relation is used).  Another, more technical approach 
to the direct TF relation is that of Willick (\cite{will94}), based on
corrections to individual galaxies within an iterative scheme.
In principle, a good knowledge of the $\log{V_{\rm M}}$ selection
function could also permit Willick's method to work for the inverse
TF relation, in order to derive the slope as it would be without 
the selection. 

Our approach necessarily cuts away a part of the sample. However,
as shown by Ekholm et al.\ (\cite{ekho98}) for the KLUN sample, the
remaining subsample suitable for the determination of $H_0$ with the
inverse TF method is still much larger than the unbiased subsample
for the direct TF relation.

The observational motivation for studying question 2) comes from   
Theureau et al.\ (\cite{theu98}), who discuss the detection rate of 21cm
line profiles at the Nan\c{c}ay radio telescope, in connection with
the angular size limited KLUN galaxy sample. 
The detection rate is relatively high, 86 percent for galaxies with
a previous redshift measurement and 61 percent for those with
unknown $z$.  However, more important than the average detection
rate, is {\em how detection depends on $p$, i.e.\ the selection function
$S(p)$}.  Unfortunately this is difficult to derive from the raw
data.  Theureau (\cite{theu97}) notes that one expects loss of galaxies
with either a narrow or broad 21cm line width.  Narrow profiles
are difficult to detect among the noise, while broad profiles,
being low, also tend to be missed.

   Hence, as a first approximation we suppose that the
distribution of $\log{V_{\rm M}}$ is affected by sharp
lower and upper cutoffs, $p_{\rm l}$
and $p_{\rm u}$, although the selection $S(p)$ more probably depends on 
various galactic parameters in a complicated manner.  
How such cutoffs influence, e.g.\ the value of $H_0$ as
determined by the inverse relation method, has not been
previously {\em quantitatively} discussed. 

We study the above questions using 
normal distributions, which allow analytic expressions.  This
forms a natural sequel to our previous discussions of the TF
relations, both direct and inverse, where the assumption of
Gaussianity has been adopted (Teerikorpi 1984, \cite{teer87},
\cite{teer90}, \cite{teer93}).
It is worth noting that though we occasionally refer to 
the KLUN sample (see e.g.\ Theureau et al.\ \cite{theua97}), the theoretical
ideas concerning the above mentioned points 1--3 have general 
validity. As an extensive sequel to the present work, 
Ekholm et al.\ (\cite{ekho98}) have
investigated whether the large value of $H_0$, derived with a
straightforward application of the inverse TF relation for the 
KLUN sample, could be due to the errors
arising from cutoff and/or calibrator sample biases. They 
show that the calibrator sample bias is more important.

In this paper we also aim at a better general understanding of the inverse
relation as a distance indicator, and present some arguments in a heuristic
manner, since the method has been justly criticised as difficult to
visualize, in comparison with the ``more natural'' direct relation approach.

This paper is structured as follows: In Sect.\ 2 we show qualitatively how
cutoffs in $\log{V_{\rm M}}$ influence the distance determination.
In Sect.\ 3 the method of normalized distance $d_{\rm n}$ is introduced
for the inverse TF relation. In Sect.\ 4, the behaviour in the $\log{H}$
vs.\ $d_{\rm n}$ diagram is calculated analytically. In Sect.\ 5, we derive
an analytical expression for the average bias in $\log{H}$ when there
is an upper or lower cutoff in $\log{V_{\rm M}}$, and give some examples
in Sect.\ 6. The fundamental problems of slope and calibration are discussed 
in Sects.\ 7 and 8. Section 9 contains concluding remarks, emphasizing the
essential points and consequences of the present study.

A note on the nomenclature: We use capital $D$'s for the diameter of a 
galaxy, $D_{25}$ is the usual apparent 25th-mag isophotal diameter
($[D_{25}]=0.1$ arcmin), $D$ is the linear diameter ($[D]=$ kpc). In 
Sect.\ 4, to make the formulae more readable, we use $\mathcal{D}$ for
the $\log{D}$'s. Small $d$'s are used for inferred distances,
$d_{\rm n}$ is the normalized and $d_{\rm kin}$ the kinematical distance.
It is assumed that kinematical distances have been calculated from a
realistic velocity field model, giving reliable relative distances
(in practice from a Virgo-centric infall model as e.g.\ in Theureau et
al.\ \cite{theub97}). The true distance is denoted by $r$.

%Section2
\section{Influence of cutoffs in $\log{V_{\rm M}}$:\ schematic treatment} 

\begin{figure}
\epsfig{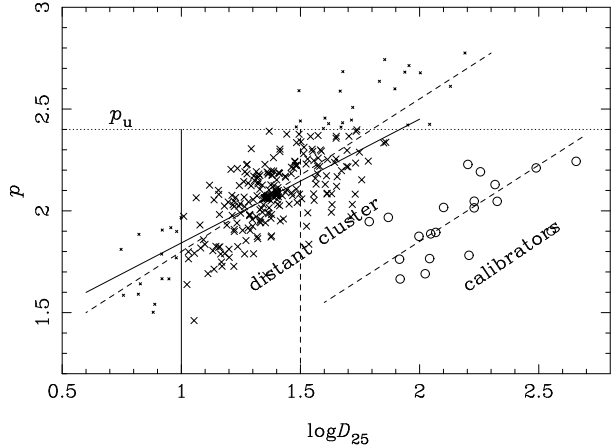}
\caption{$p$ vs.\ log of apparent diameter, $\log{D_{25}}$ diagram for
distant cluster and calibrator galaxies.
Upper cutoff at $p$ (dotted line) makes the inverse TF distance
to the cluster too small when we force the correct slope,
followed by both the calibrators and the cluster
(dashed lines), through the
cluster. The solid vertical line shows the angular diameter limit, the dashed
vertical line is the 'unbiased plateau' limit for the distant cluster. 
Cluster members that are not in the observed sample are marked with small 
symbols.}
\end{figure}

   Let us assume that the selection may be described as cutoffs on the
wings of the symmetric $p$ distribution.  To illustrate, let us consider the
simple case when there is only an upper cutoff $p_{\rm u}$, so that all $p
> p_{\rm u}$ are missed, while all $p \leq p_{\rm u}$ 
are measured.  The situation
is sketched in Fig.\ 1. The calibrator sample (at a fixed
distance) is on the right, while a distant cluster of galaxies is
on the left.  The $p_{\rm u}$-limit is indicated by a dotted line.  
The correct regression lines, with the same slopes, are shown.  Note
that unaware of the $p$ selection, one would derive from the
cluster galaxies (or more generally from field galaxies) a too shallow
slope.  For the moment we simply assume that the correct slope is
known.  We return to the problem of slope in Sect.\ 7, and
this issue will be discussed elsewhere  (Ekholm et al.\
\cite{ekho98}) in connection with real data.

In inverse TF method (iTF) for deriving distances, one basically
shifts one of the regression lines over the other, and the
required shift of $\log{D_{25}}$ gives the (logarithmic) ratio of cluster
to calibrator distance.

   In practice, one calculates for each cluster galaxy the iTF
``distance'' $d_{\rm cl}$ using the calibrating relation $p = a'\log{D} + b'$,
 so that $\log{(d_{\rm cl}/d_{\rm cal})} = \log{D(p)} - \log{D_{25}}$.  
Averaging over all cluster
galaxies gives the unbiased distance.  This is equivalent to
``shifting the regression lines''.  The iTF ``distance'' 
has a systematic error depending on $p$, though one
does not know how large the error is (c.f.\ Sect.\ 5 in \cite{T84}).

   In the depicted situation, even if one knows the
correct slope, the attempt to force-fit it on the cluster data
will shift the line towards larger $\log{D_{25}}$, because the upper cutoff
has eliminated more galaxies from above the ``correct'' (unbiased)
regression line than from below it.  Hence, the inevitable result
of an upper cutoff is that the iTF distance to the cluster will
be underestimated.

   Now, because of the limiting angular size, not all galaxies of
the cluster are included.  So one should add to the diagram a
vertical line, only galaxies to the right of which are large enough
to enter the sample. Note that this cutoff does not bias the distance
indicator if there is no $p$-cutoff.

   If one moves the cluster closer (to the right), the angular
size limit cuts away less of the cluster galaxies.  Then the
influence of the $p_{\rm u}$-cutoff on the derived distances becomes
weaker.  In terms of the calculated Hubble constant (actually up
to an unknown factor, because $b'/a'$ is still unknown), one should
see in a $\log{H}$ vs.\ $d_{\rm kin}$ diagram first at small kinematical
distances (proportional to the true ones) a constant ``plateau''
when the angular size limit keeps out of the cluster.  Then there
is an increase due to the larger influence of the $p_{\rm u}$-cutoff.

\section{``Unbiased plateau'' for the inverse relation} % 3

The plateau in the $\log{H}$ vs.\ $d_{\rm kin}$ diagram is not an
unbiased plateau; it still gives too large a value of $\langle \log{H} 
\rangle$, because of
a residual effect of the $p_{\rm u}$-limit. The bias can be reduced by
applying an upper cutoff to the cluster sample in $\log{D_{25}}$, as
illustrated by the dashed vertical line in Fig.\ 1. The galaxies
between the upper and lower $\log{D_{25}}$ cutoffs now provide the 
best iTF distance estimate.

The position of the critical righthand side vertical line in Fig.\ 1,
which cuts away the ``bad'' $\log{D_{25}}$ interval, depends on the
distance of the cluster (compare the cluster and the calibrators).
However, its position is practically constant on the {\em linear size scale},
independently of the cluster distance. This is clearly seen if one shifts in 
Fig.\ 1 the cluster galaxies to the right, i.e.\ decreases their distance
until it coincides with the distance of the calibrator cluster.
Because the TF relations are essentially the same, the critical vertical
lines will almost coincide. Hence, in the general case of a field
galaxy sample, one may shift all galaxies
independent of their distances, to one diagram of
$\log{V_{\rm M}}$ vs. $\log{D_{25}} + \log{d_{\rm kin}}$, where the abscissa 
is up to an unknown constant (as far as $H_0$ is unknown) the same as 
linear diameter, $\log{D}$.  If one now accepts
galaxies left of the (common) critical line only, an unbiased average
distance is derived.

   One might term $\log{D_{25}} + \log{d_{\rm kin}}$ 
(or $d_{\rm n} = D_{25} \cdot d_{\rm kin}$) the
normalized distance, analogously to the case
of the direct TF normalized distance introduced in \cite{T84}.  Then in
the $\log{H}$ vs.\ $\log{d_{\rm n}}$ diagram, using the correct slope $a'$, 
there is
at small normalized distances an unbiased plateau, after which
$\langle \log{H} \rangle$ starts to grow.

  The resulting unbiased plateau should be horizontal, if the slope $a'$ is
correct.  \cite{ET97} pointed out a simple device of deciding which is
the correct inverse slope:\ 
in the $\log{H}$ vs.\ $d_{\rm kin}$ diagram the run of
points should be horizontal.  This was, however, made under the
premise of no $p$-selection.  Now we understand that 
instead of $d_{\rm kin}$ one should use 
$d_{\rm n} = D_{25}d_{\rm kin}$ on the abscissa, in order to identify the 
cutoff $d_{\rm n}$. 
Only galaxies below this cutoff are expected to show a
horizontal run on the $\log{H}$ vs.\ $d_{\rm kin}$ diagram, if $a'$ 
is correct. Figure 2  shows how in the more general case, the 
lower cutoff causes
too small $\langle \log{H} \rangle$ below the unbiased plateau.
Ekholm et al.\ \cite{ekho98} will discuss the evidence for upper and
lower cutoffs in  $\log{V_{\rm M}}$ in the KLUN sample and how the 
described method can in practice be implemented.

Though in this manner one may detect sharp cutoffs in $\log{V_{\rm M}}$ and
derive an unbiased value of $H_0$ without detailed modelling, it is 
important to know how large an influence the cutoffs would have on $H_0$,
if their presence were ignored. We discuss this in Sects.\ 4--6.

\begin{figure}
\epsfig{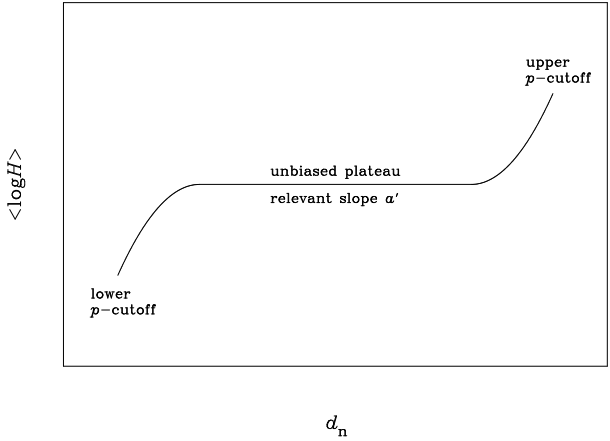}
\caption{Effect of the $p$-cutoffs on the value of $\langle \log{H} \rangle$
at different $d_{\rm n}$. The normalized distance $d_{\rm n}$ is defined
as $D_{25} d_{\rm kin}$.}
\end{figure}

\section{Analytic calculation of $\langle \log{H} \rangle$ vs.\ $d_{\rm n}$}
% 4

Assuming Gaussian distributions and sharp cutoffs
at $p_{\rm l}$ and $p_{\rm u}$, it is easy to calculate the expected run 
of $\langle \log{H} \rangle$ vs.\ normalized distance $d_{\rm n}$.  
For simplicity, we write the inverse relation as
\begin{equation}% 3
p = a'\mathcal{D} + b'
\end{equation}% 3
We use $\mathcal{D} \equiv \log{D_{\rm lin}}$ 
as shorthand notation for the logarithm of the linear diameter.
At any fixed $\mathcal{D}$ the dispersion of $p$ is $\sigma_p$.

Now the task is to calculate the bias in $\log{H}$ at each $\mathcal{D}$.  This
is done starting from the same initial formula (28) of \cite{T84} which
was used to show the absence of bias.  However, in this case
one must add the selection function $S(p)$:
\begin{equation}%{4}
\langle {\rm bias} \rangle_\mathcal{D} = \frac{\int_{-\infty}^{\infty}
(\mathcal{D}-\mathcal{D}(p))S(p)\Phi(p) dp}{\int_{-\infty}^{\infty}S(p)
\Phi(p)dp}
\end{equation}% 4
where $\Phi(p) = {\rm const} \cdot 
\exp{[-(p-p(\mathcal{D}))^2/(2\sigma_p^2)]}$ 
and $\mathcal{D}(p) = (p-b')/a'$.

   This formula gives for the sample galaxies with true (log)
diameter $\mathcal{D}$, the (log) error in the diameter deduced by the inverse
relation (using $\mathcal{D}(p)$).  It is directly reflected to the derived
value of $\langle \log{H} \rangle$ at constant $\mathcal{D}$ (i.e., at fixed 
normalized distance
${d_{\rm n}}$).

   With the present choice of the selection function (sharp upper and
lower cutoffs), and after a change of variable, 
\begin{equation}%{5}
x = \frac{p-a'\mathcal{D}-b'}{\sqrt{2}\sigma_p}, 
\end{equation}% 5
Eq.\ (4) becomes
\begin{equation}% 6
\langle {\rm bias} \rangle_\mathcal{D} = 
- \sqrt{2/\pi}(\sigma_p/a')B(x_{\rm l},x_{\rm u})
\end{equation}% 6
Here the term $B(x_{\rm l},x_{\rm u})$ can be evaluated in terms of the error
function ${\rm erf}{(x)} = (2/\sqrt{\pi})\int_0^x \exp{(-t^2)}dt$:
\begin{equation}%{7}
B(x_{\rm l},x_{\rm u}) = \frac{\exp{(-x_{\rm l}^2)}
-\exp{(-x_{\rm u}^2)}}{{\rm erf}{(x_{\rm u})}-{\rm erf}{(x_{\rm l})}}
\end{equation}% 7
where $x_{\rm l}$ and $x_{\rm u}$ correspond to the cutoffs $p_{\rm l}$ 
and $p_{\rm u}$ via Eq.\ (5).

   Inspection of the formula confirms the qualitative conclusions
of Sect.2.  Putting $x_{\rm l} = -\infty$, so that only the upper cutoff
remains (and recalling that ${\rm erf}{(-\infty)} = -1$), 
the bias is seen to be
positive: the diameters are underestimated, leading to 
underestimated distances, hence to overestimated $\log{H}$.  Note that this
calculation does not tell us the average bias in the sample.

\section{Calculation of the average bias} % 5

In order to have an idea of how large an influence the $p$-cutoffs
have on the value of $\langle \log{H} \rangle$ calculated from the whole sample
using a (correct) inverse relation, we derive the average bias
for a special case of radial space density behaviour: 
\begin{equation} % 8
n(r) = k r^{-\alpha},
\end{equation} % 8
where $r$ is the radial distance from our Galaxy and $\alpha=0$
corresponds to homogeneity.
It is easy to understand that the average bias depends on how the observed
galaxies are distributed along the $d_{\rm n}$-axis, which must depend on
how galaxies are distributed in space.

   Rather than trying to start with averaging the derived
$\langle {\rm bias} \rangle_{d_{\rm n}}$ over $d_{\rm n}$ (Eq.\ 6), we use 
another route, based on the
fact that the bivariate distribution $F(d_{\rm n},p)$ is known for an
angular diameter limited sample, when space density behaves as Eq.\ (8).
We assume for simplicity that the cosmic distribution of $p$ 
(as well as of $\mathcal{D}$) is normal, which is a rough approximation for 
fixed Hubble types, see e.g.\ Fig.\ 1 in Theureau et al.\ (\cite{theua97}).

   As illustrated by Fig.\ 1 in \cite{T84} for the case of
magnitude TF relation, in magnitude and volume
limited samples the direct regression lines are parallel and
separated by $1.382 \sigma_{M}^2$ (when $\alpha=0$).  
The separation happens so that the
volume-limited line (carrying with it the bivariate distribution)
glides along the unchanged inverse line which goes through the
average $M$ of both the volume ( $= M_0$) and magnitude limited ( $= M_0
-  1.382 \sigma_M^2$) bivariate distributions. Figure 3 gives a similar
diagram for the diameter TF relation, showing the three kinds of
regression lines for a synthetic sample of field galaxies.

\begin{figure}
\epsfig{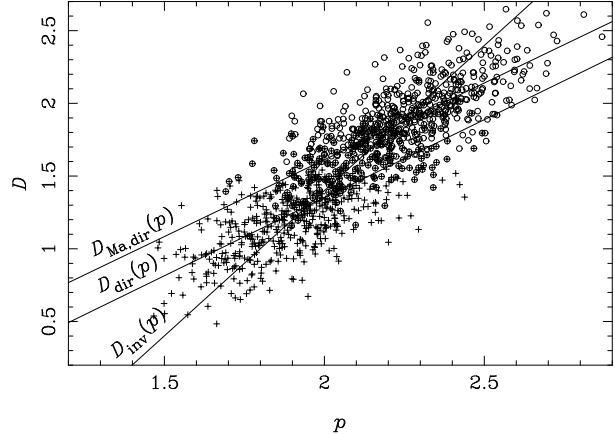}
\caption{ A volume limited subsample of a synthetic data-set is shown as
crosses, and the diameter limited subsample of the same parent data-set 
is marked with circles. Direct TF relations of these subsamples, 
$\mathcal{D}_{\rm dir}(p)$ and
$\mathcal{D}_{\rm Ma,dir}(p)$, are parallel and separated vertically by 
$(3-\alpha)\cdot \ln{10} \cdot \sigma_{\mathcal{D}_p}^2$ (Eq.\ 11). The inverse
relation $\mathcal{D}_{\rm inv}(p)$ (Eq.\ 12), that under ideal conditions 
is the same for both subsamples, intersects the direct regression lines at
$\mathcal{D}=\mathcal{D}_0$ and $\mathcal{D}=\mathcal{D}_0 
+ (3-\alpha)\cdot \ln{10} \cdot \sigma_{\mathcal{D}}^2$.} 
\end{figure}

   In the present calculation we need to know how the diameter 
direct relation is constructed from the parameters defining the iTF 
(c.f.\ \cite{T84}, Teerikorpi \cite{teer93}, where this was given for
the magnitude TF relation):
\begin{equation}%{9}
\mathcal{D}_{\rm dir}(p) = (a'/\sigma_p^2)\sigma_{\mathcal{D}_p}^2 \cdot p + 
(\mathcal{D}_0/\sigma_\mathcal{D}^2 - a'b'/\sigma_p^2)\sigma_{\mathcal{D}_p}^2
\end{equation}% 9
where
\begin{equation}%{10}
1/\sigma_{\mathcal{D}_p}^2 = 1/\sigma_\mathcal{D}^2 + 1/(\sigma_p/a')^2 .
\end{equation} % 10
The Malmquist shifted (angular size limited) direct relation is
\begin{equation}%{11}
\mathcal{D}_{\rm Ma,dir}(p) = \mathcal{D}_{\rm dir}(p) + (3-\alpha) 
\cdot \ln{10} \cdot \sigma_{\mathcal{D}_p}^2
\end{equation}%11
and the inverse relation is
\begin{equation}%{13}
\mathcal{D}_{\rm inv}(p) = (1/a')p - b'/a' .
\end{equation}% 13

   One can express the average bias at a fixed $p$ using the
inverse and (Malmquist-shifted) direct relations:
\renewcommand{\theequation}{\arabic{equation}a}
\setcounter{equation}{13}
\begin{equation}%{14a}
\langle {\rm bias} \rangle_p = \frac{\int_{-\infty}^{\infty}
(\mathcal{D}-\mathcal{D}(p))\
\Phi_{{\rm Ma,}p}(\mathcal{D})d\mathcal{D}}
{\int_{-\infty}^{\infty}\Phi_{{\rm Ma,}p}(\mathcal{D})d\mathcal{D}}
\end{equation}%14a
where 
\renewcommand{\theequation}{\arabic{equation}}
\setcounter{equation}{14}
\begin{equation}%{15}
\Phi_{{\rm Ma,}p}(\mathcal{D})=\mbox{const} \cdot 
\exp{(-(\mathcal{D}-\mathcal{D}_{\rm Ma,dir}(p))^2/
2 \sigma_{\mathcal{D}_p}^2)}
\end{equation}
so that 
\renewcommand{\theequation}{\arabic{equation}b}
\setcounter{equation}{13}
\begin{equation}%14b
\langle {\rm  bias} \rangle_p = \mathcal{D}_{\rm Ma,dir}(p) 
- \mathcal{D}_{\rm inv}(p)
\end{equation}%14b
This simple expression is then easy to integrate over all $p$ in
the sample, when one notes that the distribution of $p$ is
\renewcommand{\theequation}{\arabic{equation}}
\setcounter{equation}{15}
\begin{equation}%{16}
{\rm const} \cdot       \exp{(-(p-p_0')^2/(2 {a'}^2 \sigma_\mathcal{D}^2))}
\end{equation}%16
where $p_0'$ gives the Malmquist-shifted average value of $p$ (shifted
by $a'$ times the shift of $\mathcal{D}_0$):
\begin{equation}%{17}
p_0' = a'\mathcal{D} + b' + (3-\alpha) \cdot \ln{10} \cdot 
\sigma_{\mathcal{D}_p}^2
\end{equation}%17
Averaging $\langle {\rm bias} \rangle_p$ over all $p$ from 
$p_{\rm l}$ to $p_{\rm u}$ gives:
\begin{equation}%{18}
\langle {\rm bias} \rangle_{\rm ave}= 
-\left(\sigma_{\mathcal{D}_p}/\sigma_\mathcal{D} \right)^2 
\sigma_\mathcal{D}  \sqrt{2/\pi}B(x_{\rm l},x_{\rm u}) 
\end{equation}
where $x_{\rm l}$ and $x_{\rm u}$ are
\begin{equation}%{19}
x_{\rm l,u} = \frac{p_{\rm l,u} - p_0'}{\sqrt{2}a'\sigma_\mathcal{D}}
\end{equation}%19
Note that the information on the space density (via $\alpha$ and Eq.\ 16) is
contained in $x_{\rm l}$ and $x_{\rm u}$, which show how far away from the
Malmquist-shifted average $p_0'$ the cutoffs are, in terms of the
dispersion $\sigma_P = a'\sigma_\mathcal{D}$ of the cosmic (Gaussian) 
distribution
 function of $p$.  The factor before $B(x_{\rm l},x_{\rm u})$ can also 
be written in terms of $\sigma_p$ and $\sigma_P$.  Then
\begin{equation}%{20}
\langle {\rm bias} \rangle_{\rm ave} = - (\sigma_p^2/(\sigma_p^2+\sigma_P^2))
(\sigma_P/a')\sqrt{2/\pi}B(x_{\rm l},x_{\rm u})
\end{equation}%20
Such an analytic calculation is possible only in the given
special cases of space density.  Generally, the distribution of
points in the $p$ vs.\ $\log{d_{\rm n}}$ diagram will differ from the Gaussian
bivariate case: at each $p$, the shift of 
$\langle \mathcal{D}_{\rm dir}(p)\rangle$ 
will be due to different Malmquist biases of the 1st kind.  However, even in
this case the inverse relation is not affected and the
calculation and prediction of Sect.\ 3 is valid.

\section{Application of the average bias formula} % 6

We are interested in seeing how large an effect on $\log{H}$ 
cutoffs of different severities have.  Table 1 shows the values of
$B(-\infty,x_{\rm u})$ ($B(x_{\rm l},\infty)$) and the bias in $\langle
\log{H} \rangle$ for different values of upper 
(lower) cutoff $x_{\rm u(,l)}$, and the corresponding completeness percentage 
of the sample.

\begin{table}
\caption[]{Bias in $\langle \log{H} \rangle$ due to  
$p$ cutoffs. The upper sign in the $\pm$ symbols refers to 
$p_{\rm u}$ while the lower sign refers to $p_{\rm l}$.}
\begin{tabular}{rrcc}\hline
$x_{\rm u(,l)}$&  \% &  $-\sqrt{2/\pi}B(-\infty,x_{\rm u})$ & bias$^*$\\ 
 & & ($-\sqrt{2/\pi}B(x_{\rm l},\infty)$) & \\ \hline
$\pm \infty$ &  100 &  0.00 & 0.000\\ 
$\pm 0.84$ &  80 &  $\pm 0.22$ & $\pm 0.015$\\
0.00 &  50 &  $\pm 0.80$ & $\pm 0.056$\\
$\mp 0.84$ &  20 & $\pm 1.68$ & $\pm 0.118$ \\ \hline
\end{tabular} \\
$*$ for $\sigma_\mathcal{D}=0.28$, $\sigma_{\mathcal{D}_p}=0.14$.
\end{table}

  In particular, if half of the galaxies are lost because of the
upper cutoff (while there is no lower cutoff), formula (18)
adopts a simple form, which gives a rough estimate of the bias
for such distorted samples.
\begin{equation}%{21}
\begin{split}
\langle {\rm bias} \rangle_{50\%} & = 
-(\sigma_{\mathcal{D}_p}/\sigma_\mathcal{D})^2  
\sigma_\mathcal{D}  \sqrt{2/\pi}B(-\infty,0) \\
         & = -0.8 (\sigma_{\mathcal{D}_p}/\sigma_\mathcal{D})^2  
\sigma_\mathcal{D}  
\end{split}
\end{equation}%21
With representative values $\sigma_\mathcal{D}=0.28$ 
and $\sigma_{\mathcal{D}_p}=0.14$
this gives the bias $-0.056$, or distances underestimated by about 14 percent,
in the average. 
Intuitively, one might have expected a more substantial influence
of the $p$ cutoff, knowing that the assumption of no $p$-selection
is so fundamental for the inverse relation method.

\section{Problems of slope and calibration} % 7

   In the discussion thus far, we have assumed that the
normalized distance has been accurately calculated, so that there
are no observational errors in $\log{D_{25}}$ and the kinematical distance
scale ($d_{\rm kin}$) is accurate as well.  However, this cannot be the
case, and as a result, field galaxies collected from a fixed true
distance will show in the $p$ vs.\ $\log{d_{\rm n}}$ diagram a different slope
than without such errors.  This slope
will differ from the intrinsic slope followed by calibrators.

   Two questions arise:  first, in such a situation, what happens
to the slope-criterion requiring that 
$\langle \log{H} \rangle$ for the unbiased
plateau galaxies goes horizontally in the $\log{H}$ vs. $d_{\rm kin}$ diagram? 
Secondly, how do we calculate the actual value of $\log{H_0}$ 
with calibrators known to have a different iTF slope?  Here we
expand somewhat the compact treatment in Teerikorpi (\cite{teer90}).

   Consider two galaxy clusters in the $p$ vs.\ $\log{D_{25}}$ diagram, both
having (in their unbiased $d_{\rm n}$-part), the same iTF slope $a'$, which
however differs from the calibrator slope.  It is evident that
the slope $a'$ gives the correct relative distance of these
clusters, when one shifts the regression lines one over the other
(c.f.\ Fig.\ 1).  Hence, as an answer to the first question, this is
also the slope which corresponds to the horizontal distribution
of $\langle \log{H} \rangle$; correct distance ratios reveal the 
linear Hubble law.

   Assume now that the calibrators follow the slope $a_{\rm c}' > a'$.  One
may describe in a picturesque manner how to do the comparison
with calibrators, leading to an unbiased value of $\log{H_0}$.  As the
shallower slope for the field galaxies is assumed to be due to
errors in $d_{\rm n}$, one may think of ``adding'' random errors to
calibrator linear log diameters $\mathcal{D}$, until the slope falls 
to $a'$ and then use the
obtained relation for calibration (``shifting of regression
lines'').  Fortunately it is not necessary to actually ``shuffle''
$\mathcal{D}$'s - this is equivalent to forcing the slope $a'$ through the
calibrators, at least in an ideal situation.

   So, let us assume first that the calibrator sample is
volume-limited.  Then adding random errors
to $\mathcal{D}$ means that $\langle \mathcal{D} \rangle
 = \mathcal{D}_0$ remains the same, and one arrives at the relation
\begin{equation}%{22}
\langle p\rangle = a'\langle \mathcal{D}\rangle + b'
\end{equation}%22
On the other hand, the actual calibrators follow
\begin{equation}%{23}
\langle p\rangle = a'_{\rm c}\langle \mathcal{D}\rangle + b'_{\rm c}
\end{equation}%23
Hence, after this trick, the new zero-point which must be applied
for the sample galaxies is:
\begin{equation}%{24}
b' = b'_{\rm c} + (a'_{\rm c}-a')\langle \mathcal{D}\rangle
\end{equation}%24
There is also a longer route to this result, using Eq.\ (7) giving 
the direct relation in terms of the inverse parameters
and $\mathcal{D}_0$.  If one increases $\sigma_{\mathcal{D}_p}$ so that 
$a'_{\rm c} \rightarrow a'$ and requires that
the direct relation does not change, one also arrives at Eq.\ (24).

   On the other hand, the formula for $b'$ is exactly what is
obtained if one simply forces the slope $a'$ through the
calibrators.  From this follows that a correct calibration is
achieved by deriving the zero-point by a force-fit of the slope
$a'$ through the calibrators.

\section{The calibrator sample bias when one uses relevant field slope $a'$}

In the above discussion it is noteworthy that one does not
have to know the intrinsic slope $a'_{\rm c}$ for the calibrators.
   However, the calibrator sample must fulfill a special
condition.  It must be equivalent to a volume-limited sample, or
reflect the cosmic Gaussian distribution, as was already noted in
Teerikorpi (\cite{teer90}).  From Eq.\ (23), 
it is clear that two calibrator samples with different 
$\langle \mathcal{D}\rangle$ 
yield different zero-points $b'$ for the same $a'$ of
the field galaxies, while there can exist only one such correct
$b'$, the one corresponding to $\langle \mathcal{D}\rangle=\mathcal{D}_0$ or 
$\langle p\rangle = a'\mathcal{D}_0 + b' = p_0$.

If  $\langle \mathcal{D}\rangle$ of the calibrator sample 
deviates from the cosmic $\mathcal{D}_0$,
there is a systematical error in $b'$, resulting in an average
error in the derived $\langle \log{H} \rangle$:
\begin{equation}%{25}
\Delta \langle \log{H} \rangle = (a_{\rm c}'/a' - 1) \Delta \mathcal{D}
\end{equation}%25
or in terms of the average $p$:
\begin{equation}%{26}
\Delta \langle \log{H} \rangle = (a_{\rm c}'/a' - 1) \Delta p /a'
\end{equation}%26

   This source of error is very significant if the calibrator sample
has not been constructed with the intention of reaching the
cosmic distribution of $p$, if for example the selection of galaxies has
aimed at detection of Cepheids.  In this way, one may have a
volume-limited sample for each $p$ (c.f.\ Theureau et al.\ \cite{theua97}), but
{\em this does not guarantee that the distribution of $p$ reflects the
cosmic distribution} from which the field galaxy sample has been
drawn.

   That this could be a genuine problem, consider the calibrator sample of
KLUN with Cepheid distances. An indication of the calibrator sample bias
is seen by comparing the distribution of calibrator $p$'s with
that of the whole sample.  Because the latter is diameter
limited, it suffers from a Malmquist bias, hence its median
should be displaced to a larger value of $p$ in comparison with the
calibrators. Ekholm et al.\ (\cite{ekho98}) show that there is no 
such shift, which implies
that calibrators' $\langle p\rangle$ is larger than the cosmic value.  Such a
situation results in an overestimated $\log{H}$ when one is compelled to
use $a'<a_{\rm c}'$.

Application of Eqs.\ (25--26) requires a knowledge of the 
average $\mathcal{D}_0$
or $p_0$ of the cosmic distribution functions to be compared with
those of the calibrator sample. It is clear that any calculation 
of $\mathcal{D}_0$ 
from the field sample requires, besides a suitable method, the value
of $H_0$ itself, hence an iterative approach. Below we show that there
is another much more convenient route which does not require an 
explicit knowledge of the difference $\Delta \mathcal{D}$ 
(or the value of $H_0$). However, it 
assumes a radial space distribution law of galaxy number density.

\section{Another route:\ using calibrator slope $a'_{\rm c}$ 
and making $a'_{\rm c}/a'$ and Malmquist corrections} % 9

\begin{table}
\caption[]{$H_0$ as derived using $a'=0.5$ when $H_0({\rm true})=55$ 
and the true calibrator slope is $a_{\rm c}'$.  }
\begin{tabular}{ccc}\hline
$a_{\rm c}'$ &  $H_0({\alpha=0}) $& $H_0(\alpha = 0.8)$ \\ \hline
0.60 &  67 &      64 \\
0.65 &  74 &      68 \\
0.70 &  82 &      74 \\
0.75 &  90 &    79 \\ \hline
\end{tabular}
\end{table}

   In principle, the problem of the $p$-distribution of the
calibrators does not affect the determination of the zero-point
$b'_{\rm c}$ when one forces the correct slope $a'_{\rm c}$ 
on the calibrators. 
This makes one ask what happens if $a'_{\rm c}$ and $b'_{\rm c}$ are then
used for the field sample. First of all one expects a distance-dependent bias,
according to Eq.15 from Teerikorpi (\cite{teer90}), here adopted to diameters:
\begin{equation}%{27}
\Delta \langle \log{H} \rangle(r) =
(\langle \mathcal{D}\rangle(r)-\mathcal{D}_0)(a'_{\rm c}/a' - 1)
\end{equation}%27
Here $\langle \mathcal{D}\rangle(r)$ is the average (log) diameter 
for the galaxies at
true distance $r$. Now one may ask what is the average bias for the sample?
 The average bias clearly depends on the average of
$\langle \mathcal{D}\rangle(r)-\mathcal{D}_0$.  This is actually 
the Malmquist bias of the 1st kind for Gaussian distribution of diameters 
with dispersion $\sigma_\mathcal{D}$, and in the case of a
space density distribution $\propto r^{-\alpha}$ one gets:
\begin{equation}%{28}
\langle \Delta \log{H} \rangle = (3-\alpha) \cdot
\ln{10} \cdot \sigma_\mathcal{D}^2(a'_{\rm c}/a' - 1)
\end{equation}%28
When $\alpha=0$ the distribution is homogeneous.
In order to use this as a correction, one must know both $a'_{\rm c}$ and
$a'$.  Clearly, if the field and calibrator slopes are identical,
then $\langle \log{H} \rangle$ is unbiased, and this is so irrespective of the
nature of the calibrator sample.  
Note also that this formula contains the width 
$\sigma_\mathcal{D}$ of the total
diameter function, not the smaller dispersion 
$\sigma_{\mathcal{D}_p}$. Hence the
effect can be formidable even when the slopes $a'_{\rm c}$ and $a'$
do not differ very much.

As a concrete example, consider the KLUN field galaxies following
the inverse slope $a' \approx 0.5$ which is in agreement with
the simple disk model of Theureau et
al.\ (\cite{theua97}).  This slope, when forced through the calibrators,
gives $H_0 \approx 75 \mathrm{\,km\,s^{-1}\,Mpc^{-1}}$.  
Above it was noted that the calibrator
sample cannot represent the cosmic distribution of $\log{V_{\rm M}}$, and
this is the situation when a too high $H_0$ is derived if the actual
slope is steeper than the slope from the field galaxies.  It is
natural to ask, how large a slope $a_{\rm c}'$ would in this manner
explain the difference between $H_0({\rm dir}) \approx 55$ and 
$H_0({\rm inv}) \approx 75$. 
Taking $\sigma_\mathcal{D} \approx 0.25$, we get predicted $H_0({\rm inv})$
for different slopes $a_{\rm c}'$ as given in Table 2.

The last column corresponds to a decrease in the average space
density, corresponding to density $\propto  r^{-0.8}$ as derived
by Teerikorpi et al.\ (\cite{teer98}).  It may be concluded that in order
to explain the difference between $H_0 = 75$ and 55 solely in this
manner, it is required that actually $a_{\rm c}' \approx$ 0.65 -- 0.75, instead
of $\approx 0.5$.  Ekholm et al.\ (\cite{ekho98}) show that such
a steep calibrator slope really is consistent with the data.

\section{Concluding remarks} % 10

In the present study we have searched for answers to a few basic 
questions concerning the inverse TF relation as a distance indicator.
{\em First}, how to see the signature of observational incompleteness in the
TF parameter $\log{V_{\rm M}}$, analogously to the well-known increase in 
$\log{H}$ at increasing (normalized) distance, in the case of the direct
TF relation? {\em Secondly}, in which sense and how {\em significant} is the  
influence of the $\log{V_{\rm M}}$ incompleteness alone on the derived value
of $\log{H}$ (supposing one knows the correct inverse TF slope)?
{\em Thirdly}, how to work with the situation when the inverse slope obeyed by
the calibration sample differs from the relevant slope for the field
sample (or for a cluster), and how important is in this regard the 
nature of the calibration sample?

We have generalized the ``fine-tuning'' scheme discussed by \cite{ET97},
and have pointed out that there is a useful $\log{H}$ vs.\ $d_{\rm n}$ 
representation for the inverse relation, revealing both the relevant
inverse slope and cutoffs in  $\log{V_{\rm M}}$. Here $d_{\rm n}$
is the normalized distance applicable to the inverse TF relation,
actually the (log) linear diameter within an unknown constant.

We have derived simple analytical formulae which give the influence
of the $\log{V_{\rm M}}$ cutoffs on derived average distance (hence,
on $\log{H}$). The treatment is based on Gaussian distributions, and,
in the case of the average bias, on the assumption that the space density 
is proportional to $r^{-\alpha}$. The probable case that there 
is an upper observational cutoff in $\log{V_{\rm M}}$, would cause an
overestimated value of $H_0$. However, this effect is not at all dramatic --
even if the upper cutoff excludes half of the galaxies, $H_0$ would be 
increased typically by only 14 percent. It may be concluded that a cutoff
in $\log{V_{\rm M}}$ cannot alone explain derived large values of $H_0$.

The problems of the relevant inverse TF slope and the nature of the 
calibrator sample lead to serious consequences on the use of the inverse
TF relation as a distance indicator. It has been previously realized that
the inverse slope $a'$ applicable to the field sample is not 
necessarily the same as the slope $a_{\rm c}'$ obeyed by the calibrators,
which may result in a biased value of $H_0$ (Teerikorpi \cite{teer90}).
In the present paper, such situations were systematically studied.

If the two slopes are the same (in practice, when accuracies of calibrator 
distances are high and the photometric measurements for the field sample are
not less accurate than those for the calibrators), then there is no bias,
and it is also true that the nature of the calibrator sample (volume-limited,
magnitude-limited, etc.) is not important (\cite{ET97}).

If the sample inverse slope  $a'$ is shallower than the calibrator
slope (as normally expected), then matters come to depend critically
on the nature of the calibrator sample. First, there is the possibility
that the calibrator sample is a true, volume-limited representation of
the Gaussian cosmic distribution function of $\log{V_{\rm M}}$ from
which the field sample has been taken (the latter subject only to 
Malmquist bias). In this ideal case, it is permitted to use the field sample
slope, with the zero-point obtained by a force-fit through the 
calibrators. This was the solution suggested by Teerikorpi (\cite{teer90}),
and in this case it is not necessary to know the value of the calibrator 
slope. 

However, generally the calibrator sample's $\langle p \rangle$ is shifted
from the cosmic $p_0$. If one knows the shift $\Delta p$ (or
$\Delta \mathcal{D}$) {\em and}
the slopes $a_{\rm c}'$ and $a'$, then one may calculate the
resulting error in $\log{H}$, when the field sample's $a'$
is used. The systematic error is given by Eq.\ (25).

Another, more convenient route, 
which bypasses the explicit need for $\Delta p$, is to use
the calibrators' slope $a'_{\rm c}$ (and corresponding zero-point)
and to calculate analytically the resulting systematic error in terms
of the Malmquist bias of the 1st kind. This assumes that
the average space number density around us is proportional
to $r^{-\alpha}$. The systematic error is given by Eq.\ (27).

We repeat that in the general case both $a'_{\rm c}$ and $a'$
(with corresponding zero-points from force-fits through calibrators)
result in erroneous average distance estimates. Fortunately, we have
now some quantitative control on this error.

\acknowledgements{The major part of this paper was written when P.T.\
visited the Meudon Observatory in October 1997. He thanks Lucette Bottinelli,
Lucienne Gouguenheim and Gilles Theureau for their hospitality during the 
visit. This work has been supported by the Academy of Finland (project
``Cosmology in the local galaxy Universe''). We thank the referee for
useful criticism which helped us to clarify the presentation. We are also
grateful to Chris Flynn for helpful comments on the manuscript.}


\begin{thebibliography}{}
\bibitem[1986]{bott86}
Bottinelli, L., Gouguenheim, L., Paturel, G., Teerikorpi, P., 1986, A\&A 156, 
157
\bibitem[ET97]{ET97}
Ekholm, T., Teerikorpi, P., 1997, A\&A 325, 33 (ET97)
\bibitem[1998]{ekho98}
Ekholm, T., Teerikorpi, P., Theureau, G., Hanski, M., Paturel, G., 1998
A\&A (submitted)
\bibitem[1990]{fouq90}
Fouqu\'{e}, P., Bottinelli, L., Gouguenheim, L., Paturel, G., 1990, ApJ 349, 1
\bibitem[1994]{hend94}
Hendry, M.A., Simmons, J.F.L., 1994, Vistas Astron.\ 39, 297
\bibitem[1994]{sand94}
Sandage, A., 1994,  ApJ 430, 1
\bibitem[1995]{sand95}
Sandage, A., Tammann, G.A., Federspiel, M., 1995,  ApJ 452, 1
\bibitem[1980]{sche80}
Schecter, P., 1980, AJ 85, 801
\bibitem[T84]{T84}
Teerikorpi, P., 1984, A\&A 141, 407 (T84)
\bibitem[1987]{teer87}
Teerikorpi, P., 1987, A\&A 173, 39
\bibitem[1990]{teer90}
Teerikorpi, P., 1990, A\&A 234, 1
\bibitem[1993]{teer93}
Teerikorpi, P., 1993, A\&A 280, 443
\bibitem[1997]{teer97}
Teerikorpi, P., 1997, Ann.\ Rev.\ A\&A 35, 101
\bibitem[1998]{teer98}
Teerikorpi, P., Hanski, M., Theureau, G.\ et al., 1998, A\&A 334, 395
\bibitem[1997]{theu97}
Theureau, G., 1997, Th\`{e}se de Doctorat, (Paris Observatory)
\bibitem[1997a]{theua97}
Theureau, G., Hanski, M., Teerikorpi, P.\ et al., 1997a, A\&A 319, 435
\bibitem[1997b]{theub97}
Theureau, G., Hanski, M., Ekholm, T.\ et al., 1997b, A\&A 322, 730
\bibitem[1998]{theu98}
Theureau, G., Bottinelli, L., Coudreau, N.\ et al., 1998, A\&AS (in press)
\bibitem[1988]{tull88}
Tully, R.B., 1988, Nature 334, 209
\bibitem[1994]{will94}
Willick, J., 1994, ApJS 92, 1
\end{thebibliography}
\end{document}